\documentclass[10pt,conference]{IEEEtran}
\IEEEoverridecommandlockouts
\usepackage{cite}
\usepackage{amsmath,amssymb,amsfonts}
\usepackage{algorithmic}
\usepackage{graphicx}
\usepackage{textcomp}
\usepackage{xcolor}
\usepackage{subcaption}
\usepackage{booktabs}
\usepackage{caption}
\usepackage{pgfplots}
\usepackage[english]{babel}
\usepackage{csquotes}
\usepackage{comment}
\usepackage{tabularx}
\usepackage{todonotes}
\usepackage{placeins}

\usepackage{hyperref}
\hypersetup{colorlinks=true, urlcolor=blue}
\usepackage{cleveref}

\def\BibTeX{{\rm B\kern-.05em{\sc i\kern-.025em b}\kern-.08em
    T\kern-.1667em\lower.7ex\hbox{E}\kern-.125emX}}
\begin{document}

\title{Distributed Quantum Circuit Optimisation: Evaluating Global and Local encodings
\thanks{M.G.G. acknowledges support from the EPSRC UK Quantum Technologies Programme under grant EP/T001062/1
and VeriQloud.
M.H. acknowledges support from the Business Finland through project \textit{SeQuSoS} (Grants No. 112/31/2024), 
and Research Council of Finland through project \textit{Profi 8 - qSIME} (Grants No. 365343).}}

\author{\IEEEauthorblockN{1\textsuperscript{st} Maria Gragera Garces}
\IEEEauthorblockA{\textit{Quantum Software Lab, University of Edinburgh} \\
Scotland, UK \\
0009-0000-9018-7435}
\and
\IEEEauthorblockN{2\textsuperscript{nd} Majid Haghparast}
\IEEEauthorblockA{\textit{Faculty of Information Technology, University of Jyväskylä} \\
Jyväskylä, Finland \\
0000-0003-3427-5961}
}

\maketitle

\begin{abstract}
As distributed quantum architectures begin to emerge, understanding the interaction between quantum circuit optimisation and circuit partitioning becomes increasingly important. In this work, we study how circuit optimisation influences distributed quantum workloads under system-level trade-offs. We compare three compilation strategies (global optimisation, local optimisation, and a hybrid approach) across a large benchmark suite of quantum algorithms. Using telegate-based partitioning, we evaluate the resulting distributed circuits in terms of gate counts, circuit depth, the number of induced non-local gates, and compilation overhead, thereby approximating computational, communication, and classical preprocessing costs.

Our results show that circuit optimisation does not uniformly benefit distributed execution. Local optimisation minimises computational resources and achieves the lowest compilation overhead among the optimising strategies. Global optimisation can reduce communication cost even though it is not explicitly communication-aware. The hybrid strategy can simultaneously reduce both computational and communication overhead, but at the expense of significantly increased compilation time.
\end{abstract}

\begin{IEEEkeywords}
Quantum computing, Distributed Quantum Computing, Circuit Optimisation
\end{IEEEkeywords}

\vspace{-0.3 cm} \section{Introduction}
Distributed quantum computing (DQC) represents a promising path toward scalable quantum computation. 
In this paradigm, multiple quantum processing units are interconnected through quantum networks, allowing circuits that exceed the qubit capacity of individual devices to be partitioned and executed across physically separated processors.

Distributing these quantum circuits introduces substantial overhead. 
Both computational cost (such as circuit depth and multi-qubit gate count) and communication cost (non-local operations requiring entanglement) are expensive and noise-sensitive in near-term architectures. 
Reducing the circuit-level cost footprint of distributed workloads is therefore an essential step toward long-term scalability.

For the past three decades \cite{barenco1995elementary}, 
circuit optimisation has played a central role in minimising inefficiencies in quantum computations. 
Optimisation techniques reduce gate counts and circuit depth by exploiting algebraic equivalences and commutation relations between quantum operations \cite{voichick2023cognac, nam2018automated}. 
However, exact quantum circuit optimisation is computationally hard \cite{van2023optimising,kjelstrom2025exact}. 
Even in heuristic and non-deterministic settings, 
optimisation runtime can be substantial \cite{zilk2025breaking},
making it an significant bottleneck for future large computations \cite{quetschlich2023reducing}.

Traditional optimisation strategies operate on complete, monolithic circuits and leverage global circuit structure.
Distributed optimisation methods designed to maximise parallel computation \cite{cuomo2023optimized} 
and minimise communication overhead \cite{Mengoni2025efficient} 
have been explored in the literature. 
In both works, circuit rewriting is performed prior to partitioning, 
thereby optimising the monolithic circuit before distribution.
In contrast, local optimisation 
(applied independently to distributed subcircuits) 
has received no systematic study. 
The global/local encoding distinction was first introduced in the context of distributed error mitigation~\cite{GrageraGarces2026dqc_zne}; 
in this work we ask how the same structural question applies to circuit optimisation.

We explicitly compare these two paradigms, as well as a hybrid strategy that applies both global and local optimisation, 
namely pre- and post- circuit partitioning optimisations, to understand how the ordering of optimisation and partitioning within the compilation pipeline impacts distributed quantum workloads.
All strategies are evaluated against a distributed, non-optimised baseline. 
Our goal is to determine whether globally informed or locally constrained optimisation is better suited for distributed quantum execution.

Concretely, we conduct a systematic empirical study across an extensive benchmark suite, 
evaluating three compilation strategies under circuit-level cost metrics including 
single-, two-, and three-qubit gate counts, circuit depth, the number of induced non-local gates,
and classical compilation overhead.
Our results reveal a structural trade-off between computational compression and communication cost, 
demonstrating that the stage at which optimisation is applied significantly influences distributed workload efficiency. 
We further show that this trade-off can be partially mitigated by a hybrid strategy, which simultaneously reduces computational and communication cost, 
albeit at the expense of increased compilation overhead. 
Efficient compilation is particularly critical in distributed and parallel settings \cite{czarnul2025optimization}, 
and the additional overhead introduced by hybrid optimisation should therefore be viewed as a substantive system-level trade-off.\

To explore optimisation strategies for DQC, 
we must first establish two key enabling technologies: 
circuit partitioning
and circuit optimisation. 

\vspace{-0.1 cm} \subsection{Circuit Partitioning}

In order to distribute a quantum circuit that is too large to be implemented within a single device, 
we must partition it into smaller subcircuits that can be executed across multiple collaborating devices.
A common abstraction enabling this distribution is the dependency hypergraph \cite{andres2019automated}. 
Dependency hypergraphs are constructed by mapping each qubit to a vertex and each multi-qubit gate to a hyperedge connecting the corresponding vertices.
Under a fully interconnected network setting, where each device has equal capacity (i.e., can host the same number of qubits),
general \textsc{Balanced Hypergraph Partitioning} (BHP) solvers with adaptive capacity penalties
are sufficient to partition circuits such that the resulting subcircuits fit within device constraints
while minimizing cross-QPU communication.

Cross-QPU communication arises from gates (hyperedges) that span multiple partitions. 
These cross-partition operations can be implemented using either classical or quantum communication channels \cite{barral2025review}.
Classical communication methods, most notably quasi-probability decomposition–based circuit cutting \cite{peng2020simulating}, 
incur exponential sampling overhead \cite{jing2025circuit}, 
as they require repeated execution of subcircuits to probabilistically reconstruct non-local operations.
In contrast, quantum communication primitives rely on the consumption of pre-shared EPR pairs to implement teleportation-based interactions.
In both settings, communication is costly and time-intensive and should therefore be minimized. 
Moreover, it degrades output quality: in the quantum setting through additional physical noise, 
and in the approximate classical setting through increased estimator variance and reduced confidence in reconstructed fidelities \cite{chen2022approximate}.

\vspace{-0.1 cm} \subsection{Circuit Optimisation}
Quantum circuit optimisation has emerged as a critical technique for improving the performance of quantum computers given an input set of commands. 
By exploiting gate identities and equivalences, optimisation passes can reduce circuit depth, gate count, and overall noise accumulation without altering the implemented unitary operation. 
Optimisation techniques include: 
gate cancellation (removes redundant operations) \cite{nam2018automated}, 
circuit resynthesis (reconstructs circuit segments using fewer gates) \cite{lau2025lazy}, 
and topology-aware routing (minimizes the overhead of mapping logical circuits to physical devices) \cite{davis2020towards}. 

Previous literature explores optimisation techniques for DQC, 
with a particular focus on either parallelism, the time it takes to complete the computation in a distributed setting compared to a non-distributed (monolithic) setting \cite{cuomo2023optimized}, 
or the minimization of non-local gates consuming quantum network resources via communication primitive optimizations \cite{Mengoni2025efficient,newopt} such as burst communication \cite{wu2022autocomm, andres2024distributing}.
More broadly, end-to-end compilation pipelines that jointly address partitioning, remote operation scheduling, and local subcircuit optimisation have been proposed for DQC~\cite{Ferrari2023modular}.

\vspace{-0.1 cm} \section{Methodology}\label{sec:method}

To evaluate the impact of optimisation encodings within DQC compilation,
we compare three encodings (global, local, and hybrid)
across a benchmark of quantum circuits,
using existing partitioning and optimisation strategies.

\vspace{-0.1 cm} \subsection{Benchmark Circuits}

We compare our encodings on the non-optimised MQT Bench benchmark suite \cite{quetschlich2023mqt} algorithms, 
implemented with Qiskit \cite{fingerhuth2018open}. 
The benchmarking suite includes the implementation of key quantum algorithms as well as random circuits accross 2-130 qubits. 
While random circuits have been shown to distort circuit partitioning results \cite{grageragarces2026distortion}, they remain well-established baselines for evaluating circuit optimisation, which is the primary focus of this work.
We believe the 1943 
circuits tested constitute a substantial and representative benchmark set of quantum circuit workloads.

\vspace{-0.1 cm} \subsection{Optimisation Encodings}
The tested encodings include: 

\begin{itemize}
\item \textit{Local Encoding:} The circuit is first partitioned into subcircuits, then each subcircuit is independently optimized. 
This approach allows optimisation to target the specific structure of each subcircuit after distribution boundaries are established; 
consequently it does not consider cross-partition interactions during optimisation.
\item \textit{Global Encoding:} The full circuit is first optimized globally, then partitioned into subcircuits. This approach preserves global optimisation opportunities before distribution constraints are introduced.
\item \textit{Hybrid Encoding:} The circuit is optimized both before and after partitioning.
\item \textit{Only Distribution (Baseline):} The circuit is partitioned and distributed without any optimisation. This isolates the effects of distribution alone and demonstrates the necessity of optimisation in the distributed setting.
\end{itemize}

\vspace{-0.1 cm} \subsection{Experimental Setup} 

We use KaHyPar \cite{schlag2023high}, a state-of-the-art solver for the Balanced Hypergraph Partitioning (BHP) problem,
to partition circuits into subcircuits.

All experiments assume a fully interconnected quantum network.
For each circuit, we evaluate configurations ranging from 2 to 10 QPUs,
where the total network qubit capacity equals the original circuit width.
That capacity is partitioned across devices (with up to 3\% imbalance tolerance),
and no additional qubit overhead is introduced for communication.
Reported results are averaged over all nine network sizes.

Circuit optimisation employed Qiskit's built-in transpiler pass manager, 
configured at optimization level 3, 
which applies heavy optimisation via 
commutative gate cancellation, 
resynthesis of two-qubit blocks via Cartan's KAK decomposition, 
and measurement-oriented unitarity-breaking passes \cite{ibmquantum_setoptimization}.

We measured three primary metrics:
\begin{itemize}
\item Computational cost: the depth of the resulting sub-circuits, as well as the number of three- , two- and single- qubit gates present in these subcircuits.
\item Communication cost: the number of non-local gates that must be implemented to enable cross-QPU collaboration (in other words the number of partitioned hyperedges by the BHP solver). 
We note that this metric abstracts over network topology and gate-level entanglement cost: a fully interconnected network assumption ignores the effect of path length on entanglement sharing \cite{BurtChenLeung2025entanglement}, 
and different non-local gate types may consume varying numbers of EPR pairs. 
Despite this, non-local gate count provides a tractable, hardware-agnostic proxy for communication overhead.
\item Runtime: Total execution time from circuit submission to result retrieval, providing insight into the computational efficiency of each encoding.
\end{itemize}

\vspace{-0.2 cm} \section{Results} \label{sec:results}
\vspace{-0.1 cm}
\subsection{Computational optimisation}
\begin{figure}[t]
        \centering
        \begin{subfigure}{0.2\textwidth}
            \centering
            \includegraphics[width=\linewidth]{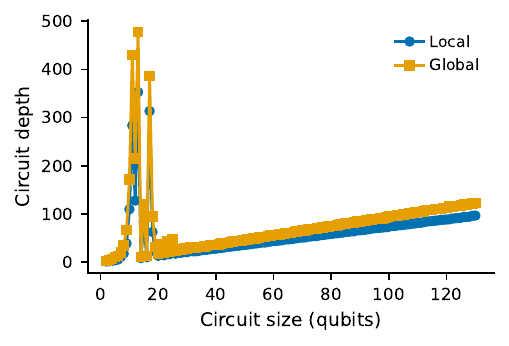}
            \caption{Mean distributed circuit depth.}
            \label{fig:depth}
        \end{subfigure}
        \begin{subfigure}{0.2\textwidth}
            \centering
            \includegraphics[width=\linewidth]{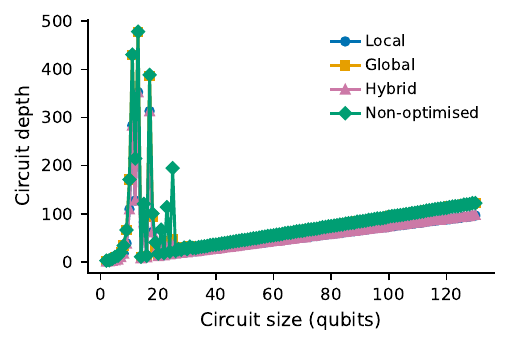}
            \caption{Mean depth including baselines.}
            \label{fig:depthbaselines}
        \end{subfigure}
        \begin{subfigure}{0.2\textwidth}
            \centering
            \includegraphics[width=\linewidth]{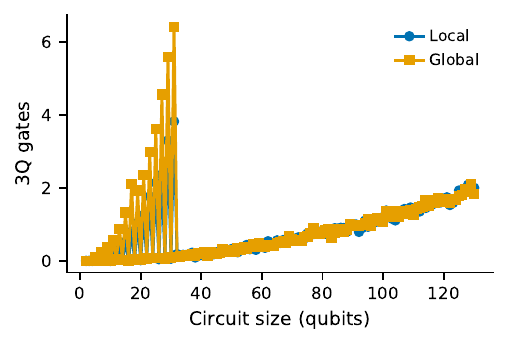}
            \caption{Mean three-qubit gate count.}
            \label{fig:3q}
        \end{subfigure}
        \begin{subfigure}{0.2\textwidth}
            \centering
            \includegraphics[width=\linewidth]{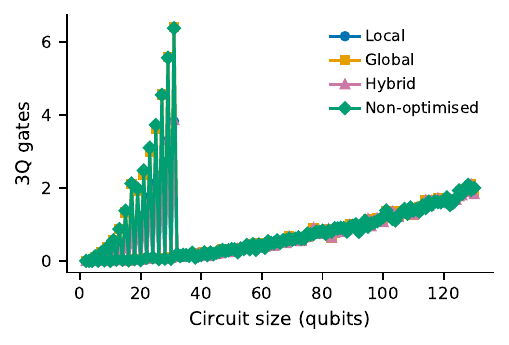}
            \caption{Three-qubit gates including baselines.}
            \label{fig:3qbaselines}
        \end{subfigure}
        \begin{subfigure}{0.2\textwidth}
            \centering
            \includegraphics[width=\linewidth]{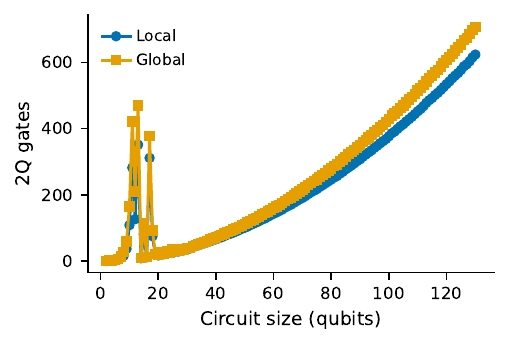}
            \caption{Mean two-qubit gate count.}
            \label{fig:2q}
        \end{subfigure}
        \begin{subfigure}{0.2\textwidth}
            \centering
            \includegraphics[width=\linewidth]{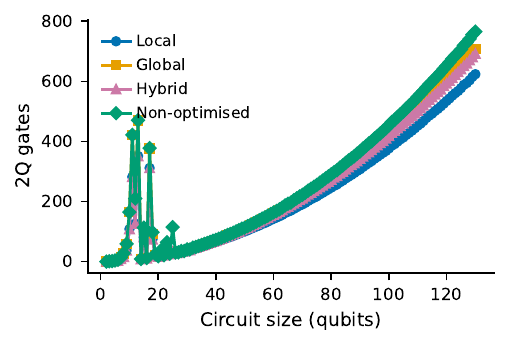}
            \caption{Two-qubit gates including baselines.}
            \label{fig:2qbaselines}
        \end{subfigure}
        \begin{subfigure}{0.2\textwidth}
            \centering
            \includegraphics[width=\linewidth]{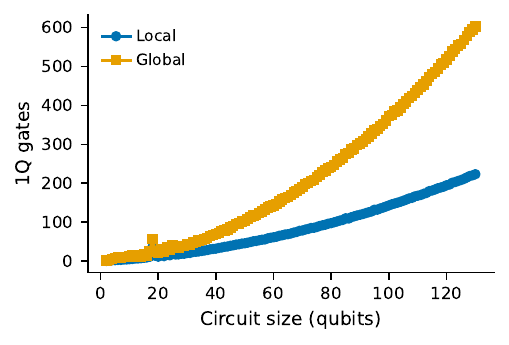}
            \caption{Mean single-qubit gate count.}
            \label{fig:1q}
        \end{subfigure}
        \begin{subfigure}{0.2\textwidth}
            \centering
            \includegraphics[width=\linewidth]{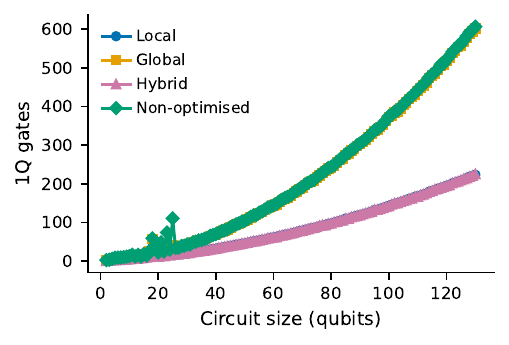}
            \caption{Single-qubit gates including baselines.}
            \label{fig:1qbaselines}
        \end{subfigure}
        \caption{Computational resource comparison across optimisation encoding strategies after distribution.
        Global shown in yellow, Local in blue, Hybrid in pink and Non-optimised in green.}
        \label{fig:bigfigcomp}
        \vspace{-0.1 cm}
    \end{figure}

We first analyse the impact of optimisation strategy on computational resources after distribution.
Figure~\ref{fig:bigfigcomp} reports the average single-qubit (1Q), two-qubit (2Q), and three-qubit (3Q) gate counts, as well as circuit depth.

\paragraph{Single-qubit gates}
Local and hybrid encodings produce the lowest 1Q counts (performing equally).
Global optimisation slightly improves over the non-optimised baseline, but remains significantly worse than other strategies.

\paragraph{Two-qubit gates}
Local encodings yield the lowest 2Q count.
Hybrid and global strategies are comparable, with hybrid being marginally better at scale.
Both outperform the non-optimised baseline.

\paragraph{Three-qubit gates}
Hybrid encoding achieves the lowest 3Q count.
Local and global encodings are comparable, with global performing slightly worse at larger scales.
All optimised strategies improve over the non-optimised baseline.

\paragraph{Circuit depth}
Local and hybrid encodings achieve the lowest depth and scale.
Global optimisation provides only marginal improvement over the non-optimised baseline.

Overall, local and hybrid optimisations consistently 
and comparably reduce computational cost and depth, 
while global encodings provide limited benefits when compared to 
non-optimised implementations.
    
\vspace{-0.1 cm} \subsection{Communication cost reductions}
\label{subsec:comm}

\begin{figure}
        \centering
        \begin{subfigure}{0.2\textwidth}
            \centering
            \includegraphics[width=\linewidth]{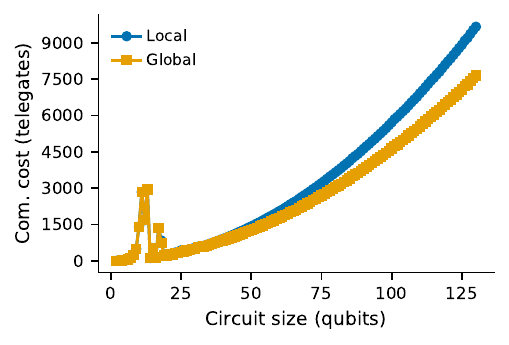}
            \caption{Mean communication cost.}
            \label{fig:depth}
        \end{subfigure}
        \begin{subfigure}{0.2\textwidth}
            \centering
            \includegraphics[width=\linewidth]{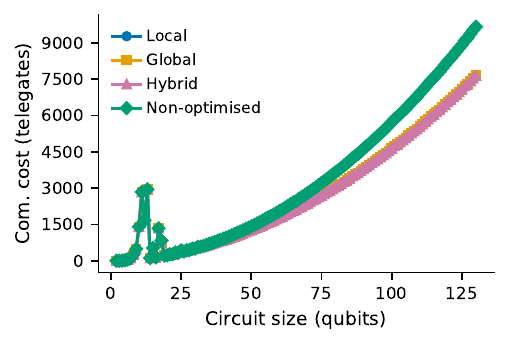}
            \caption{Mean communication cost including baselines.}
            \label{fig:depthbaselines}
        \end{subfigure}
        \caption{Mean inter-QPU communication cost under different optimisation encodings.    }
        \label{fig:comm}
    \vspace{-0.3 cm}
\end{figure}

Next we analyse the impact of optimisation encoding on inter-QPU communication cost.
Figure~\ref{fig:comm} reports the mean communication cost after partitioning.

Up to approximately 40 qubits, all strategies exhibit comparable communication overhead.
Beyond this scale, global and hybrid encodings increasingly reduce communication cost relative to both the non-optimised and local strategies.
The reduction becomes more pronounced asymptotically, indicating that optimisation prior to or across partitions can restructure circuits to limit cross-partition interactions at scale.

\vspace{-0.1 cm} \subsection{Compilation time per strategy}
\label{subsec:runtime}

We measure the compilation time required by each optimisation strategy.
Table~\ref{tab:comp_time} reports the mean, median, and standard deviation across all benchmarks.

As expected, the non-optimised baseline is the fastest, as it avoids optimisation and repartitioning overhead.

Local encoding incurs modest overhead and remains substantially faster than both global and hybrid strategies.

Global and hybrid encodings exhibit mean runtimes, 
with hybrid encodings costing almost double the time of global when looking at median compilation time.

\begin{table}
    \centering
    \scriptsize
    \begin{tabular}{c|c|c|c}
    Encoding & Mean time & Median time & Standard Deviation time \\
    \midrule
    Local & 0.124146 & 0.0482 & 0.869361 \\
    Global & 0.208103 & 0.04835 & 1.178517 \\
    Hybrid & 0.29163 & 0.0833 & 1.719969 \\
    No optimization & 0.04221 & 0.0108 & 0.082127 \\
    \end{tabular}
    \caption{Compilation time per optimisation strategy (seconds).
    Hybrid encoding exhibits the highest variance and median runtime, while local encoding remains comparatively efficient.}
    \label{tab:comp_time}
    \vspace{-0.5 cm}
\end{table}

\vspace{-0.3 cm} 
\section{Discussion}
\label{sec:disc}

Our results reveal a trade-off between computational optimisation, communication reduction, and compilation time.

Local encoding consistently minimises gate counts and circuit depth while maintaining low compilation overhead.
Hybrid encoding achieves comparable computational improvements, with asymptotically higher two-qubit counts.
Global encoding provide limited to no optimization. 
This behaviour likely arises because global optimisation operates over a substantially larger search space.
Exact quantum circuit optimisation is computationally intractable in general, 
with circuit equivalence and redundancy detection problems known to be QMA-complete \cite{janzing2005non}, 
and routing and synthesis tasks NP-hard even in restricted settings \cite{sivarajah2021t}.
As circuit size increases and partitions are introduced, optimisation opportunities become diluted, reducing the effectiveness of global transformations.

In contrast, global and hybrid encodings reduce inter-QPU communication at larger circuit scales, 
whereas local encoding offers no improvement over the non-optimised baseline.
This is expected, as local optimisation is applied after partitioning and therefore does not fundamentally alter the cut structure.
The communication advantage of global and hybrid strategies becomes increasingly pronounced as circuit size grows. 
Even when not explicitly communication-aware, reducing multi-qubit interactions prior to partitioning can lead to fewer cross-QPU dependencies.

These communication gains, however, come at the cost of higher compilation time and greater runtime variance, particularly for the hybrid strategy.
Compilation overhead becomes increasingly relevant at scale, where optimisation time may rival or exceed execution latency in distributed settings.
Hybrid encoding offers the most balanced improvement in computation and communication, at the cost of the largest compilation overhead, with median compilation time nearly doubling that of the other strategies.

No single encoding dominates across all metrics.
Local optimisation favours computational efficiency and compiler speed, whereas global strategies favour communication reduction at scale.
Hybrid encoding are the best of both worlds but come with high compilation costs.
The appropriate optimisation strategy therefore depends on whether the dominant bottleneck lies in computation, communication, or compilation latency.

Future work should include evaluating a broader range of optimisers and partitioners, as the current results may be influenced by the specific combination used in this work.
A more extensive benchmark suite, particularly incorporating workloads beyond the circuit model, would also be beneficial.
Additionally, exploring the impact of distribution-aware partitioning strategies \cite{cuomo2023optimized,Mengoni2025efficient} and topology-aware distribution \cite{BurtChenLeung2025entanglement}, in this setting, remains an open direction.
Following from this, another important question is how to choose among these strategies under differing constraints, such as quantum network topology and workload size.

Code and data available: \href{https://github.com/grageragarces/DQopt}{github.com/grageragarces/DQopt}.

\vspace{- 0.2 cm}
\bibliographystyle{IEEEtran}
\bibliography{references}

\end{document}